\PassOptionsToPackage{dvipsnames}{xcolor}

\documentclass[acmsmall,authorversion,nonacm]{acmart}

\AtBeginDocument{%
	}

\setcopyright{acmcopyright}
\copyrightyear{2023}
\acmYear{2023}
\acmDOI{XXXXXXX.XXXXXXX}

\acmJournal{JACM}

\author{Dominique Geissler}
\affiliation{%
  \institution{LMU Munich \& Munich Center for Machine Learning}
  \city{Munich}
  \country{Germany}}
\email{d.geissler@lmu.de}
 
\author{Stefan Feuerriegel}
\affiliation{%
  \institution{LMU Munich \& Munich Center for Machine Learning}
  \city{Munich}
  \country{Germany}}
\email{feuerriegel@lmu.de}

\title{Analyzing the Strategy of Propaganda using Inverse Reinforcement Learning: Evidence from the 2022 Russian Invasion of Ukraine}

\usepackage{booktabs}
\usepackage{siunitx}
\usepackage{multirow}
\usepackage{makecell}
\usepackage[
    colorinlistoftodos,
    textsize=footnotesize,
        ]{todonotes}

\definecolor{darkgreen}{rgb}{0.0, 0.5, 0.0}

\usepackage{xspace}
\usepackage{enumitem}
\usepackage{url}
\usepackage{caption}
\usepackage{subcaption}
\usepackage[user,titleref]{zref}
\begin{document}

\begin{abstract}
The 2022 Russian invasion of Ukraine was accompanied by a large-scale, pro-Russian propaganda campaign on social media. However, the strategy behind the dissemination of propaganda has remained unclear, particularly how the online discourse was strategically shaped by the propagandists' community. Here, we analyze the strategy of the Twitter community using an inverse reinforcement learning (IRL) approach. Specifically, IRL allows us to model online behavior as a Markov decision process, where the goal is to infer the underlying reward structure that guides propagandists when interacting with users with a supporting or opposing stance toward the invasion. Thereby, we aim to understand empirically whether and how between-user interactions are strategically used to promote the proliferation of Russian propaganda. For this, we leverage a large-scale dataset with 349,455 posts with pro-Russian propaganda from 132,131 users. We show that bots and humans follow a different strategy: bots respond predominantly to pro-invasion messages, suggesting that they seek to drive virality; while messages indicating opposition primarily elicit responses from humans, suggesting that they tend to engage in critical discussions. To the best of our knowledge, this is the first study analyzing the strategy behind propaganda from the 2022 Russian invasion of Ukraine through the lens of IRL.
\end{abstract}

\maketitle


\section{Introduction}

Information warfare has emerged as a new phenomenon whereby social media is increasingly used to gain a competitive advantage over opponents \cite{Thornton.2015}. As part of information warfare, propaganda is strategically used to manipulate public opinion and shape narratives around wars \cite{Bail.2020, Golovchenko.2018, Ratkiewicz.2018}. Propaganda is defined as a \emph{systematic} and \emph{coordinated} effort to spread and amplify manipulative content through social interactions \cite{Jowett.Chapter1.2012}. Hence, social media poses vulnerabilities that propaganda can exploit, as it enables discourse between broad audiences, which can strategically be shaped by online communities.

There is repeated evidence that social media such as Twitter have been used for information warfare and, specifically, for the dissemination of propaganda \cite{Alieva.2022, Badawy.2018, Doroshenko.2021}. For example, a Russian organization known as the Internet Research Agency (IRA) was found to spread propaganda on Twitter \cite{Bail.2020, Doroshenko.2021, Luceri.2020}. Recent evidence has documented that the 2022 Russian invasion of Ukraine was also accompanied by a large-scale propaganda campaign. Here, previous research has provided evidence of coordinated efforts to shape discourses on Facebook and Twitter \cite{Geissler.2022, Pierri.2022}. Russian propaganda aims to promote hostility against the West and undermine support for Ukraine among the international community \cite{Golovchenko.2018, BBC.2022, Scott.2022}. As a result, propaganda may destabilize countries by polarizing citizens and increasing political division \cite{Yablokov.2022, Miller.2019}.

\textbf{Research question (RQ):} This research seeks to analyze the dissemination strategy of Russian propaganda. Specifically, we aim to understand how propagandists interact with other users to shape online discourse dynamics and to promote the proliferation of Russian propaganda, especially in the context of support and opposition in online communities. To this end, we study the following research question: \emph{How do propagandists strategically shape the online discourse?} 

Russian propaganda spreaders strategically shape the dynamics of the online discourse on online platforms such as Twitter through coordinated actions. Members of the coordinated network interact with each other to push their content and build on each other. However, propagandists are not only confronted with supporters but also with opponents in online communities. While some members of the international community amplify Russian propaganda, others oppose it by countering propaganda. For example, previous research found that pro-Russian disinformation around the annexation of Crimea was accompanied by a larger wave of disagreement \cite{Golovchenko.2020}. Hence, when shaping online discourses, propagandists are likely to interact with users that have a supporting or opposing stance towards the invasion. Nevertheless, little is known about the strategy of propagandists to shape online discourse dynamics, particularly when interacting with supporting and opposing replies.

By analyzing the interactions between propagandists and other users, we gain insights into the strategy that is used to amplify specific narratives and influence public opinion. The study of these coordinated actions allows us to unveil the underlying mechanisms through which propaganda spreads, evolves, and gains traction within the digital realm. Such insights are essential for developing effective countermeasures to mitigate the influence of propaganda campaigns. Moreover, examining the impact of Russian propaganda spreaders on online discourses provides valuable insights into the vulnerabilities of online platforms. By shedding light on the strategy used to manipulate public opinion, we can identify the weaknesses in online platforms that are susceptible to information warfare. This understanding is crucial for policymakers, social media platforms, and researchers to design robust defenses, policies, and technological solutions that safeguard the integrity of online discourse and protect democratic processes.

In this paper, we analyze the strategy of Russian propaganda to shape the online discourse on Twitter, particularly when interacting with users with a supporting or opposing stance toward the invasion. Specifically, we aim to understand empirically whether and how between-user interactions are strategically used to shape the online discourse. For this, we leverage inverse reinforcement learning (IRL). Specifically, IRL provides a principled approach to model a user's online behavior over time (i.e., the sequence of interactions with supporters and opponents) as a Markov decision process (MDP). IRL then allows us to infer the otherwise unseen reward structure that guides the online behavior of propagandists when shaping the online discourse. To account for potential differences in online characteristics between bots and humans, we analyze the strategy by both separately.

We build upon a large dataset of Russian propaganda messages from Twitter \cite{Geissler.2022}.\footnote{With propaganda messages we refer to the different types of tweets possible on Twitter: source tweets, retweets, replies, and mentions.} Our dataset includes 349,455 propaganda messages that were posted by 132,131 accounts between February through July 2022. Moreover, we analyze 13,919 replies and mentions to propaganda messages that were classified into support and opposition. To the best of our knowledge, this is the first study to leverage IRL to analyze the strategy of propagandists when shaping the online discourse, particularly in the context of support and opposition. 

\textbf{Contributions:}\footnote{Codes and data for reproducibility will be available in a public repository upon publication.} 
\begin{enumerate}
    \item We propose the use of inverse reinforcement learning to analyze the interaction strategy of propagandists when shaping online discourse. 
    \item We find that propagandists strategically make use of resharing and posting organic content to shape the dynamics of the online discourse. In addition, users rely on the resharing behavior of others to disseminate propaganda.
    \item We confirm empirically that bots and humans pursue different strategies when it comes to shaping the online discourse: While bots and humans strategically make use of others to reshare their content, humans make strategic use of opposition in their replies and tend to engage in critical discussion. In contrast, bots act regardless of support and opposition in replies and mentions.
\end{enumerate}

\section{Related work}

\textbf{Propaganda:} Propaganda has already been used in ancient times \cite{Jowett.Chapter2.2012}, while it is nowadays used for domestic and foreign influence. As an example of the former, domestic media in Russia was forced to adopt the official narrative around the 2022 invasion by new domestic legislation \cite{Alyukov.2022, Sloane.2022}. Nowadays, propaganda is extensively shared through online channels due to their interaction-based structure. For example, recent work found that propaganda is spread online through coordinated communities \cite{Hristakieva.2022}. Different from this work, previous research aimed at online propaganda was mostly focused on propaganda detection, specifically through natural language processing \cite{DaSanMartino.2019, Rashkin.2017, Maarouf.2023}.

\textbf{Russian propaganda on social media:} Russian propaganda has been repeatedly disseminated through social media due to the low cost of sharing large volumes of propagandistic content and the unregulated nature of social media platforms \cite{Ferrara.2016, Im.2020, Miller.2019}. On the one hand, Russian propaganda aims at influencing domestic affairs, e.g., with regard to the Russian opposition leader Alexei Navalny \cite{Alieva.2022}. On the other hand, Russian influences on foreign affairs have been documented. For example, the Internet Research Agency (IRA) was suspected to have meddled with foreign political events such as the 2016 U.S. presidential election \cite{Badawy.2018, Shao.2018}, the U.K. Brexit Referendum \cite{Grcar.2017}, and the 2017 French presidential election \cite{Ferrara.2017}. The U.S. Congress even launched an official investigation of the interference of the IRA during the 2016 U.S. presidential election \cite{USSenate.2019}, and, as a result, Twitter aimed to find and remove accounts that are associated with the IRA.\footnote{\url{https://blog.twitter.com/official/en us/topics/company/2018/2016-election-update.html}} 

Over time, propaganda campaigns have become more advanced. For example, fake profiles that pretended to be genuine users were created to spread pro-Russian messages during the beginning of the 2022 invasion of Ukraine \cite{BBC.2022}. \citet{Pierri.2022} found that Russian propaganda and low-credibility content was shared on Facebook and Twitter through superspreaders, while \citet{Geissler.2022} found that Russian propaganda was disproportionately spread by bots. However, the strategy behind the campaign has remained largely unknown. Motivated by this, we provide new insight by analyzing the strategy of propagandists during the 2022 Russian invasion of Ukraine.

\textbf{Online coordination and herding behavior:} Misinformation and propaganda have been spread on social media in a coordinated manner, inter alia through the use of bots \cite{Ferrara.2016, Geissler.2022}. Online coordination can be measured with the Synchronized Action Framework \cite{Magelinski.2022, Ng.2022}: The framework measures coordination as (1)~semantic coordination (the use of common hashtags), (2)~referral coordination (the use of common URLs), and (3)~social coordination (the use of common mentions). Previous research found coordination efforts on social media during the Russo-Ukrainian war \cite{Geissler.2022}. This research adds by analyzing the strategy behind the coordinated propaganda campaign. 

Online crowds that are confronted with coordinated misinformation or propaganda can employ collective intelligence to identify false information \cite{Levy.1997}. However, collective intelligence can be undermined by herding behaviors, where individuals imitate the behavior of their peers \cite{Lorenz.2011}. For example, one work has modeled the spreading dynamics of true vs. false rumors and thereby finds that virality is driven by herding tendencies (and that it is not reduced by collective intelligence) \cite{Prollochs.2023}.

\textbf{Methods for modeling online behavior:} Researchers have developed various methods to study online behavior. One literature stream aims to analyze how users help in making content go viral. For example, research has studied whether users tend to reshare emotional speech faster than neutral speech \cite{Stieglitz.2013}, spread true or false information \cite{Vosoughi.2018}, or make hatespeech go viral \cite{Maarouf.2022}. However, the analyses in this stream are typically made at the message level and not at the user level. Different from the approach in our paper, this stream cannot capture between-user interactions as it considers the virality of messages through the help of users rather than the individual behavior of users.

A different literature stream uses machine learning (e.g., tailored deep neural networks) to model propagation paths of social media \cite[e.g.,][]{Ratkiewicz.2018}. However, while such models can typically capture a rich set of variables and account for interactions among users, such models typically act as black box. Therefore, such models do not aid in our research questions to \emph{understand} the strategy of propaganda dissemination.

\textbf{Research gap:} Little is known about the strategy with which propagandists shape the discourse dynamics on social media. In this paper, we close this gap by leveraging IRL to empirically analyze the strategy of propagandists in the dissemination of propaganda, specifically in the context of support and opposition.

\section{Data}

Our analysis is based on a large-scale dataset of Russian propaganda messages from Twitter \cite{Geissler.2022}. The dataset includes 132,131 users that have posted 349,455 messages from February through July 2022. We filtered the dataset for English-speaking users who have posted at least five messages and received at least five retweets or replies from other users. This leaves 1054 users that posted 38,267 messages and received $\sim$12,700 retweets. Russian propaganda reached $\sim$350,000 users, which we measure as the number of unique followers of authors in our dataset \cite{Cha.2010}. The dataset contains messages that include pro-Russian propaganda hashtags such as \texttt{\#istandwithrussia}, \texttt{\#standwithrussia}, \texttt{\#istandwithputin}, \texttt{\#standwithputin}. Further, we crawled replies and mentions ($N =$ 13,919) that respond to these messages using the Twitter API with the twarc2 mentions\footnote{\url{https://twarc-project.readthedocs.io/en/latest/twarc2_en_us/\#mentions}} and conversations archive\footnote{\url{https://twarc-project.readthedocs.io/en/latest/twarc2_en_us/\#conversations}} endpoints. We classified the replies and mentions into supporting and not-supporting the messages they refer to.

\section{Methods}
\label{sec:methods}

\begin{figure}[t]
    \centering
    \includegraphics[width=\textwidth]{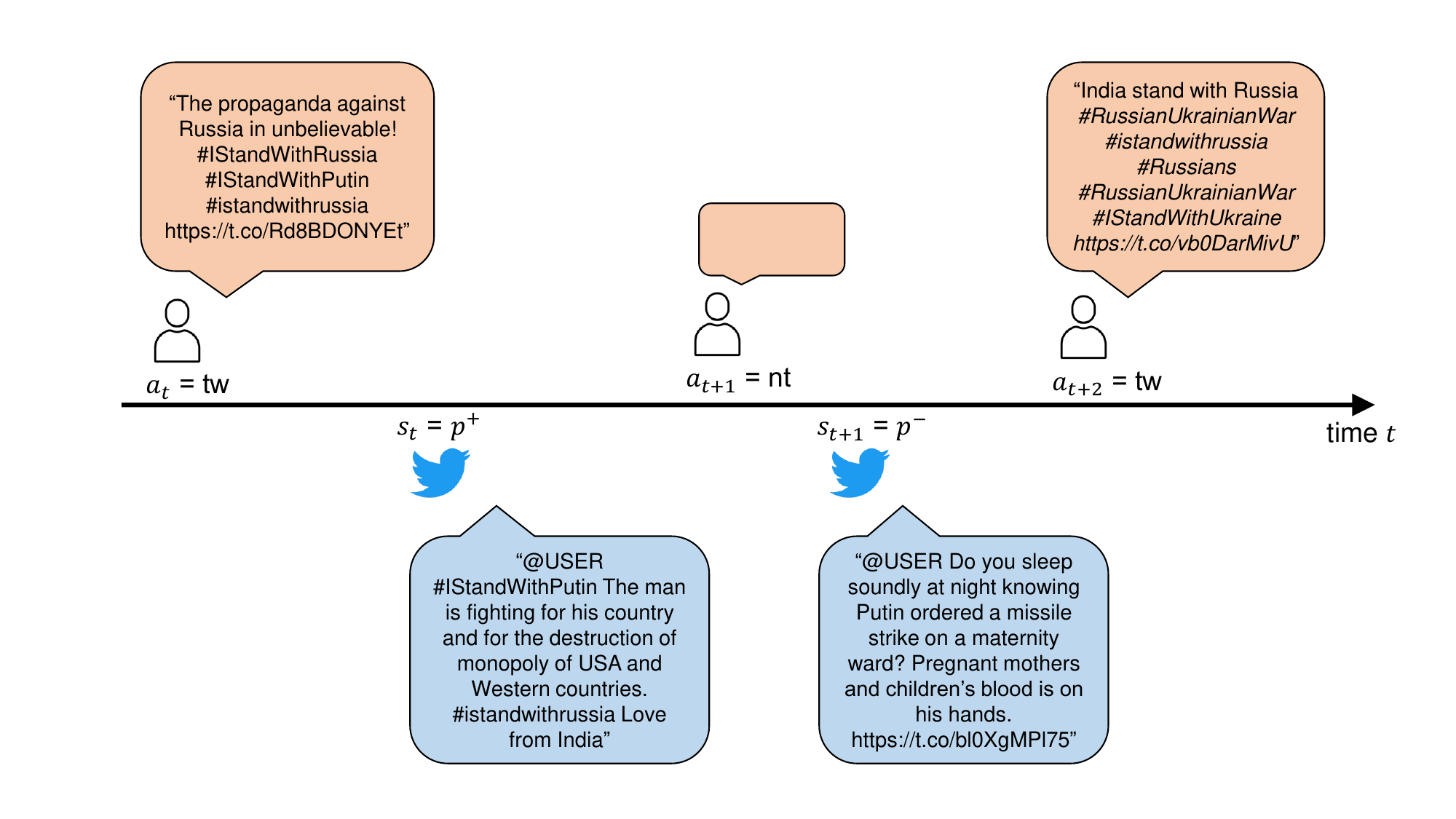}
    \caption{Example of a trajectory that occurred in the dataset.}
    \label{fig:trajectory_example}
\end{figure}

In this section, we first explain our methods for detecting bots and humans. Next, we classify replies and mentions into their stance towards the invasion, which provides the basis for parts of our subsequent analyses. Using this information, we develop our MDP to model user interactions on social media. An example of a trajectory of interactions can be found in Figure~\ref{fig:trajectory_example}. Lastly, we explain how we estimate the rewards in our MDP using IRL and thereby infer the strategy of users to shape the online discourse. 

\subsection{Objectives}

In this work, we seek to analyze the strategy of propagandists when shaping the dynamics of the online discourse on Twitter, particularly in the context of support and opposition.  We aim to analyze empirically whether and how between-user interactions are systematically used to proliferate online propaganda. For this, we model social media behavior as a MDP where a user interacts with other users through messages. In our MDP, the rewards capture which form of interaction is used by users in which situation (e.g., whether a user tends to respond to a retweet by posting new content or by posting a reply), yet the rewards are unknown and must be estimated. Specifically, we leverage IRL to infer the rewards structure that guides the propagandists' strategy. The reward structure thus captures the strategy with which propagandists employ online interactions to shape discourse dynamics. Analyzing the strategy of propagandists must carefully account for the interactions between users over time  (e.g., an opposing message in response to propaganda may make it likely that other users do not comment, while maybe a supporting response may elicit further responses in the long run). Hence, this requires a \emph{sequential} model such as a MDP. In contrast, descriptive analyses are not sequential but static, and thus cannot capture complex dependencies in user behavior, because of which they are limited in the extent to which they can characterize strategies.

\subsection{Bot detection} 

The users in our dataset were classified as bots vs. humans using Botometer \cite{Varol.2017}. Botometer is a supervised machine learning classifier that assesses the likelihood of an account being bot or human based on account features, the friendship network, and linguistic features of the account's message history. Following earlier research \cite{Shao.2018, Wojcik.2018}, we classified accounts with a Botometer score $>0.5$ as bots. Overall, 642 accounts with bot information had posted at least 5 messages and had received at least 5 retweets or replies. Of those accounts, 11.05\% are classified as bots and 88.95\% as humans. Later, we also report results from validation with an alternative algorithm for detecting bots, which eventually led to conclusive findings regarding the propaganda strategy of bots and humans.

\subsection{Stance detection}

We crawled English replies and mentions in addition to the messages from \cite{Geissler.2022}. We then computed the average stance of the replies and mentions toward original pro-Russian messages using BERTweet \cite{Nguyen.2020}. Thereby, we account for the fact that replies are not necessarily supporting the stance of the message they are referring to but that a subsequent discussion may involve user interactions that both show support or opposition to the original message. 

Inspired by prior research \cite{Bar.2023}, we followed a three-step procedure. (1)~We labeled a sample of 1250 replies and mentions by their stance using crowd workers (\url{www.mturk.com}). Specifically, we asked three workers to state whether a reply (a)~supports, (b)~opposes, or (c)~is neutral towards the message it refers to. We take the majority vote as the aggregated label (if all answers were different, we assumed the reply is neutral). (2)~We mapped the previous label onto a binary label. We thus assigned the label ``support'' if the crowdworkers' label is (a)~support; and we assign the label ``opposition`` if the crowdworkers' label is either (b) or (c). (3)~We next computed the most important tokens that express support/opposition (via log odds ratio) \cite{Kawintiranon.2021}. We added them to the BERTweet vocabulary. This should allow our BERTweet model to capture vocabulary that is unique to the Russo-Ukrainian war. (4)~We fine-tuned BERTweet on the 1250 stance labeled data and computed the average support/opposition stance for the remaining replies and mentions. We map the average support/opposition stance to a binary scale with a threshold of 0.5. Hence, replies and mentions with an average stance > 0.5 were classified as support and vice versa. Later, we perform robustness checks for the threshold.

\begin{figure}[t]
    \centering
    \vspace{-0.4cm}
    \includegraphics[width=0.7\textwidth]{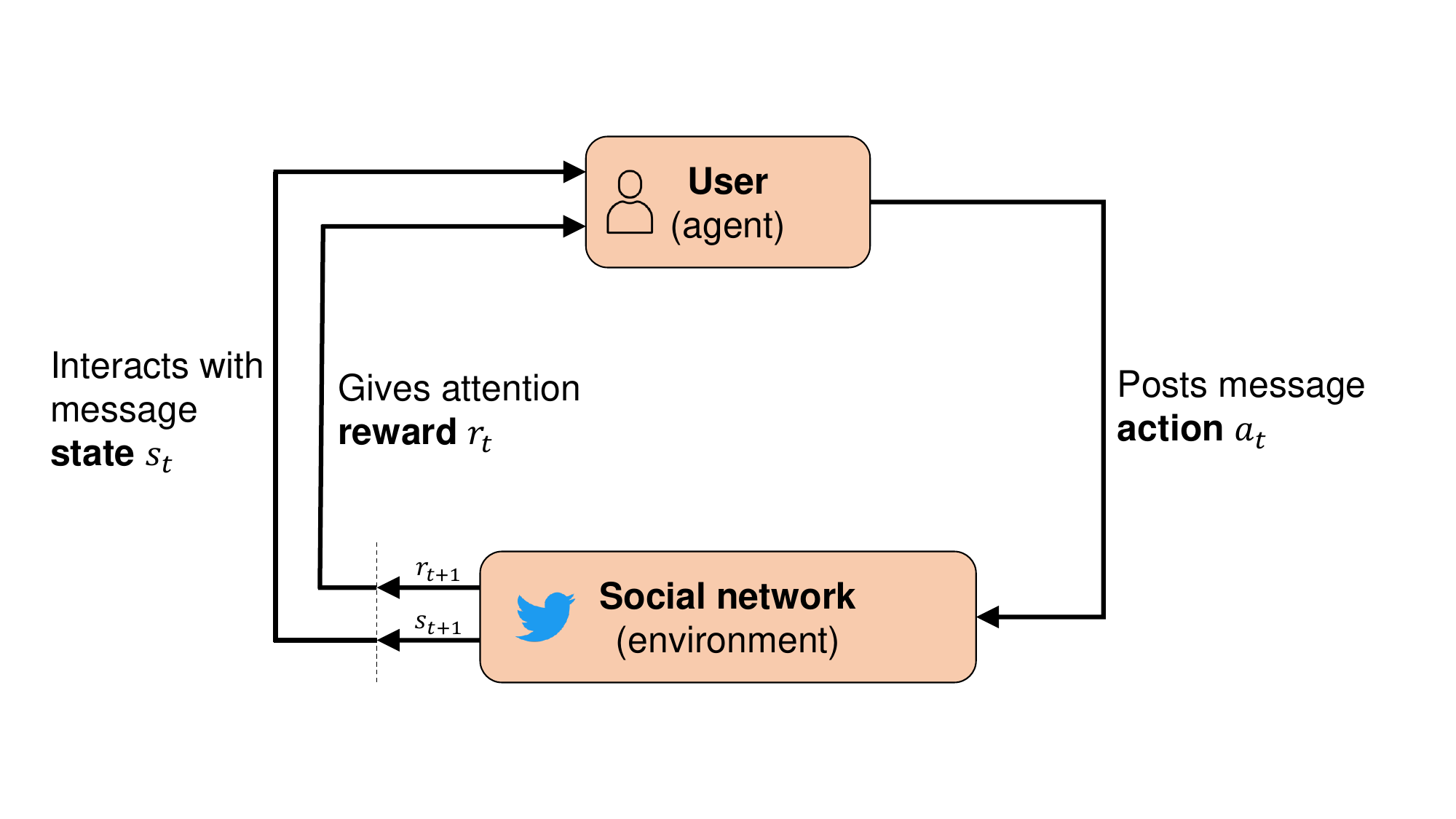}
    \vspace{-1cm}
    \caption{Overview of our MDP modeling online interactions.}
    \label{fig:twitter_mdp}
\end{figure}

\subsection{Model development}
To infer the propaganda strategy of bots and humans, we model user interactions via a MDP. A MDP is given by a 4-tuple $(\mathcal{S}, \mathcal{A}, \mathcal{T}_a, \mathcal{R}_a)$ consisting of (1)~a state space $\mathcal{S}$, (2)~an action space $\mathcal{A}$, (3)~a transition probability $\mathcal{T}_a:\mathcal{S} \times \mathcal{S} \to [0, 1]$, and (4)~a reward function $\mathcal{R}_a: \mathcal{S} \times \mathcal{S} \mapsto \mathbb{R}$ \cite{Sutton.2018}. Analogous to earlier research \cite{Das.2014, Luceri.2020}, MDPs allow us to model online interactions over time. Specifically, we can capture a social media user (\emph{agent}), who interacts with other users online (\emph{environment}) through messages (\emph{actions}). Note that, for each user, we use and estimate a separate MDP. Figure~\ref{fig:twitter_mdp} shows an overview of our MDP. The specifications of our MDP are as follows.

\textbf{Actions:}  A user can make one of the following actions $a \in \mathcal{A}$ given by$\mathcal{A} \in \{ \mathsf{tw}, \mathsf{rt}, \mathsf{rp}, \mathsf{nt}\}$ with: 
\begin{itemize}
    \item \emph{tweet ($\mathsf{tw}$)}: the user generates new content by posting a source tweet,
    \item \emph{retweet ($\mathsf{rt}$)}: the user shares existing content from others,
    \item \emph{reply ($\mathsf{rp}$)}: the user interacts with content by replying to or mentioning others, and
    \item \emph{nothing ($\mathsf{nt}$)}: the user does not perform an action.
\end{itemize}

\noindent
\textbf{States:} Recall that we model a specific user who can interact with other users, where the latter is represented by the environment. In our MDP, we capture the responses of other users through the {states}. To this end, we define the state space $\mathcal{S} = \{ \mathsf{t}, \mathsf{p^+}, \mathsf{p^-}, \mathsf{n} \}$ with:
\begin{itemize}
    \item \emph{retweet ($\mathsf{t}$)}: the messages of the agent (e.g., an source tweet, retweet, or reply) are shared by a user from the social network,
    \item \emph{supportive reply ($\mathsf{p^+}$)}:  the messages of the agent receive a supportive response in form of a reply or mention by a user from the social network,
    \item \emph{opposing reply ($\mathsf{p^-}$)}: the messages of the agent receive a opposing response in form of a reply or mention by a user from the social network, and
    \item \emph{nothing ($\mathsf{n}$)}: none of the users from the social network engage with the messages of the agent.
\end{itemize}
Summarizing, actions refer to the behavior that a user does actively, while states refer to the behavior of other users and are thus exogenous. 

\textbf{Transitions:} The above formalization allows us to model interactions over time. At each time step $t = 1, 2, \ldots$ (measured in seconds), the user makes an interaction (\emph{action}) $a_t \in \mathcal{A}$ based on current {state} $s_t \in \mathcal{S}$, that is, based on the responses of the social network. As such, users from the social network may respond to the {action} through, e.g., a retweet or a reply, denoted by the {state} $s_{t+1} \in \mathcal{S}$. For example, let us consider a user who posts a message and has thus performed action $a_t = \mathsf{tw}$. Subsequently, the user receives a supporting reply ($s_t = \mathsf{p^+}$). Before the user reacts, the user received another reply, this time it is in opposition ($s_{t+1} = \mathsf{p^-}$). As a result, the user transitions from {state} $s_t = \mathsf{p^+}$ to $s_{t+1} = \mathsf{p^-}$. The probability of moving from one {state} $s_t$ to a new {state}  $s_{t+1}$ due to some action $a_t$ is given by the {transition probability} $\mathcal{T}_a$, i.e., $P_{a}(s_{t+1} = s' \mid s_{t} = s, a_t = a)$. Figure~\ref{fig:trajectory_example} shows an overview of the example trajectory and the content of the messages.

For notation, let us denote a series of transitions by \emph{trajectory} $\zeta =\{ \langle s_1, a_1 \rangle, \langle s_2, a_2 \rangle, \langle s_3, a_3  \rangle, \ldots \}$. Hence, the online behavior of a user $i$ is thus captured by the trajectory $\zeta_i$. For this, we order the online activities (\emph{actions}) of user $i$ (\emph{agent}) and the interactions from the social network (\emph{states}) chronologically. At each time step $t$, the user is in exactly one {state} and takes exactly one {action}. The user is in {state} $s_t = \mathsf{n}$ if no other user from the social network interacted with the user $i$ (agent). For example, the user posts a series of new messages before the other users had time to interact with the new content. Likewise, a user performs {action} $a_t = \mathsf{nt}$ if the user does not react to other users, i.e., the other users interact with the content of the agent several times before she/he has time to react.

\textbf{Rewards:} Each transition from state $s_t$ to $s_{t+1}$  generates an immediate {reward} $r_t = \mathcal{R}_{a_t}(s_t, s_{t+1})$ due to action $a_t$. Importantly, the reward structure encodes the strategy with which a user promotes the dissemination of propaganda content. For example, if the reward is larger for a retweet than for a reply in a given state, this implies that the user generally prefers to retweet as part of her/his dissemination strategy. Formally, users follow some policy that maximizes their reward over time. Here, the cumulative reward is given by $\mathbb{E}[\sum_{t=1}^\infty \gamma^t \mathcal{R}(s_t, s_{t+1})]$, where $\gamma \in (0,1)$ is a discount factor, which determines the importance of short-term and long-term {rewards}. In our work, the {rewards} are unknown, and we thus estimate them later through the use of IRL.

\subsection{Inverse reinforcement learning}

We leverage IRL to capture the dissemination strategy of bots and humans.\footnote{IRL is related to reinforcement learning (RL), yet the objective behind both is different. In RL, an agent learns to navigate through an uncertain environment with trial-and-error actions \cite{Kaelbling.1996}. The correct actions are indicated through a reward function. Here, the rewards are \emph{known}, and the objective is to estimate the optimal policy (actions). This is different from IRL, where the reward function is \emph{unknown} and needs to be estimated through observations from some behavioral policy.} The goal of IRL is to estimate the reward function $\mathcal{R}_a$ from the observed behavior by an {agent}. In our MDP, this means that we recover the unknown rewards of the user $i$ from her/his trajectory $\zeta_i$. Specifically, IRL recovers the reward structure of each \emph{state-action} pair $\langle s,a \rangle$ by finding a function $g(s,a)$ that matches $\mathcal{R}_a$ in our MDP. Recall that we model each user $i$ and her/his interactions as a separate MDP. Hence, we get one reward function for each user. Later, we can then compare the distribution of the rewards by bots against the distribution of rewards by humans.

The use of IRL has several benefits for our study. To this end, we can infer and eventually imitate the behavior of users (\emph{agent}), even when the reward function is not explicitly given \cite{Ng.2000, Ramachandran.2007}. For this reason, IRL has been previously used to understand human and animal behavior in behavioral studies \cite{Das.2014, Touretzky.1997, Watkins.1989}. 
In our work, knowing the underlying strategy and thus being able to imitate propaganda dissemination on social media can eventually aid in curbing the spread of propaganda campaigns on social media.

\textbf{Estimation procedure:} We use maximum entropy IRL \cite{Ziebart.2008} to estimate the reward function $g$. Maximum entropy IRL is a popular approach to IRL as it solves a common problem in IRL, namely, that the {state-action} space can be too large to estimate a {reward} for each {state-action} pair \cite{Abbeel.2004, Ng.2000}. Instead, maximum entropy IRL approximates the {state-action} space by mapping a set of features $f$ to the reward function $\mathcal{R}$ as a weighted linear combination of the feature values, i.e., $g(f) = g(f,\theta) = \theta^T f = \mathcal{R}$. 

In this work, the user is in exactly one \emph{state} and performs exactly one \emph{action} at each time step $t$. Hence, there are 16 possible combinations of \emph{state-action} pairs. We represent each {state-action} pair by the vector $(\mathsf{t}, \mathsf{p^+}, \mathsf{p^-}, \mathsf{tw}, \mathsf{rt}, \mathsf{rp})^T$. The former three values represent the {states} while the latter three represent the {actions}. Similar to a one-hot encoding, the values are binary variables, which are $=1$ depending on the {state} and the {action} of the user. The state $s = \mathsf{n}$ and the action $a = \mathsf{nt}$ are represented by setting the values corresponding to the {states} and {actions} to 0, respectively. For example, if the users from the social network have retweeted user $i$ ($s_{t} = \mathsf{t}$) but she/he does not react by not taking a new action ($a_t = \mathsf{nt}$), then the vector is given by $(1, 0, 0, 0, 0, 0)^T$. This allows us to represent the state-action space by a $6 \times 16$ feature matrix $f$, where each state-action pair is represented by a unique vector.

The reward function is represented by a 16-dimensional vector, where each element represents the {reward} for a certain {state-action} pair. A high reward in a {state-action} pair indicates that the user is likely to perform {action} $a$ in {state} $s$ and vice versa. Similarly, a low reward indicates that a user is less likely to perform {action} $a$ in {state} $s$. 
Then, our IRL algorithm takes as input the feature matrix $f$, the trajectory $\zeta$, and the transition probabilities $\mathcal{T}$ for each user and estimates their reward functions.

\textbf{Interpretability:} 
Maximum entropy IRL uses a linear model $\mathcal{R} = \theta^T f$. This means that we can infer the weights of the features in $f$ by solving the equation for $\theta$ using the known features in $f$ and the estimated rewards $\mathcal{R}$ from the maximum entropy IRL. This gives us an estimate of the importance of each feature in $f$. Again, we compute the feature importance for each user separately, so that we can directly compare bots vs. humans.

\textbf{Implementation details:}  In our work, we use the implementation of maximum entropy IRL from \cite{Luceri.2020}.
As in their work, the trajectory $\zeta_i$ for each user $i$ must have a minimum length of five, meaning there are at least five {actions} and at least five responses represented by {states} for reasons of numerical stability during estimation. In addition, we set the discount factor $\gamma = 0.9$.

\subsection{Ethical statement}
This research did not involve interventions with human subjects, and, thus, no approval from the Institutional Review Board was required by the author institutions. All analyses are based on publicly available data, and we do not make any attempt to track users across different platforms. We comply with the privacy policy of Twitter, which states users agree that their content can be viewed ``by the general public'' unless settings are set otherwise.\footnote{\url{https://twitter.com/en/privacy}} We respect the privacy of users by not publishing usernames in our paper and only report aggregate results.

\section{Results}

\subsection{Stance of replies and mentions}

Russian propaganda spreaders receive responses from the social network in the form of replies and mentions. These replies and mentions can support ($s = \mathsf{p^+}$) or oppose ($s = \mathsf{p^-}$) pro-Russian messages. We classified replies and mentions into support and opposition using our above BERT-based stance detection. As a result, the ratio of supportive vs. opposing replies and mentions was 40.66\% vs. 59.34\% for samples labeled by crowdworkers and 25.55\% vs. 75.45\% for samples labeled by our classifier. Examples of supportive and opposing messages can be found in Table~\ref{tbl:example_replies}. We performed a qualitative study of the underlying content. When the replies and mentions are supportive, they tend to pick up the pro-Russian and/or anti-Western theme of the original message. In contrast, opposing replies and mentions counter the original messages, e.g., by pointing toward the crime of invasion.

\begin{table}
	\scriptsize
	\begin{tabular}{p{1cm}p{6cm}p{6cm}}
		\toprule
		 \textbf{Stance} & \textbf{Source tweet} & \textbf{Reply (example)} \\
		\midrule
		\multirow{8}{*}{Supportive} & \#IStandWithPutin God bless Russia, God bless Putin.. https://t.co/PRjCgKS1pP & @USER \#Putin is totally Right \#IStandWithPutin  \\ 
            \cline{2-3}
             & The propaganda against Russia in unbelievable! \#IStandWithRussia \#IStandWithPutin \#istandwithrussia https://t.co/Rd8BDONYEt & @USER \#IStandWithPutin The man is fighting for his country and for the destruction of monopoly of USA and Western countries.  \#istandwithrussia Love from India \\
             \cline{2-3}
             & America has always played tricks. This time he is getting a befitting reply. India is always with Russia. \#IStandWithPutin https://t.co/FYQ2YDzTjx & @USER Not only India we Africans \#IStandWithPutin \\
            \midrule 
		\multirow{8}{*}{Opposing} & We Support Rus We Support Putin decision And India also support Russia \#StandWithRussia \#StandWithPutin \#SupportRussia @USER @USER https://t.co/o3ykllG730 & @USER @USER @USER Then you support terrorists bombing innocent people. You are as good as terrorist yourself https://t.co/fzgqwd18Kx \\
            \cline{2-3}
            & Protecting your country is not a guilt.  \#istandwithrussia \#IStandWithPutin https://t.co/R6BrGgC7a0 & @USER Invading a neighboring country unprovoked is a crime. \\
            \cline{2-3}
            & \#IStandWithPutin \#ISupportPutin \#iSupportRussia Western media is portraying PUTIN as a monster but the real monsters are in West... https://t.co/7I0XNjstCa & @USER Russians are killing children. Putin is the monster. https://t.co/giXs53kwxl \\           
		\bottomrule
	\end{tabular}
	\caption{Examples of supportive and opposing replies to pro-Russian messages. We replace the mentioned usernames with ``@USER'' to protect the identity of users.}
	\label{tbl:example_replies}
\end{table}

\subsection{Strategy to shape online discourse dynamics}

\begin{figure*}[htp]
  \centering
  \vspace{-0.4cm}
  \begin{minipage}{.54\textwidth}
  \centering
      \includegraphics[width=\textwidth]{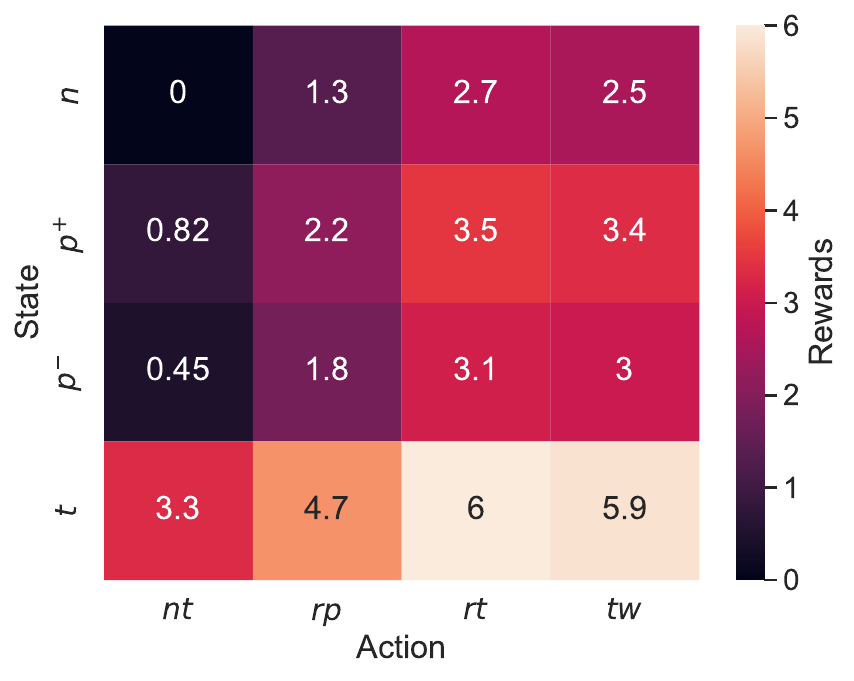}
      \caption{Average estimated rewards for {state-action} pairs for propagandists.}
      \label{fig:heatmap_all}
  \end{minipage}\hfill
    \begin{minipage}{.45\textwidth}
    \centering
    \includegraphics[width=\textwidth]{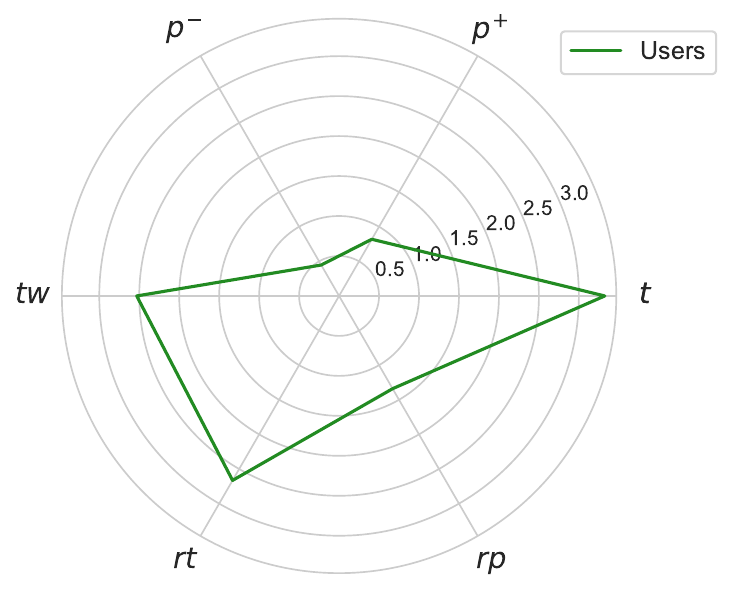}
    \caption{Weights of the reward function in our IRL algorithm. Shown is a breakdown by the states and actions for propagandists.}
    \label{fig:spiderchart_all}
    \end{minipage}
\end{figure*}

We now analyze the reward structure in our MDP. The reward structure represents the strategy with which between-user interactions are strategically used by propagandists to shape the dynamics of the online discourse. Figure~\ref{fig:heatmap_all} shows the average rewards for each state-action pair for propagandists. We find that propagandists strategically reshare others and post new content to shape the online discourse. In addition, the users rely on the resharing behavior of others to spread propaganda. Overall, propagandists appear to be foremost incentivized to take any action after having been reshared. 

To quantify the importance of the states and actions, we further compute the average weight of each state and action. Figure~\ref{fig:spiderchart_all} corroborates our above findings: users make strategic use of posting new content, resharing others, and being reshared by others. The importance of users' replies as well as receiving support or opposition are minor.

\subsection{Strategy of bots and humans}
\begin{figure}[t]
    \centering
      \vspace{-0.4cm}
          \begin{subfigure}[t]{0.49\textwidth}
            \centering
            \includegraphics[width=\textwidth]{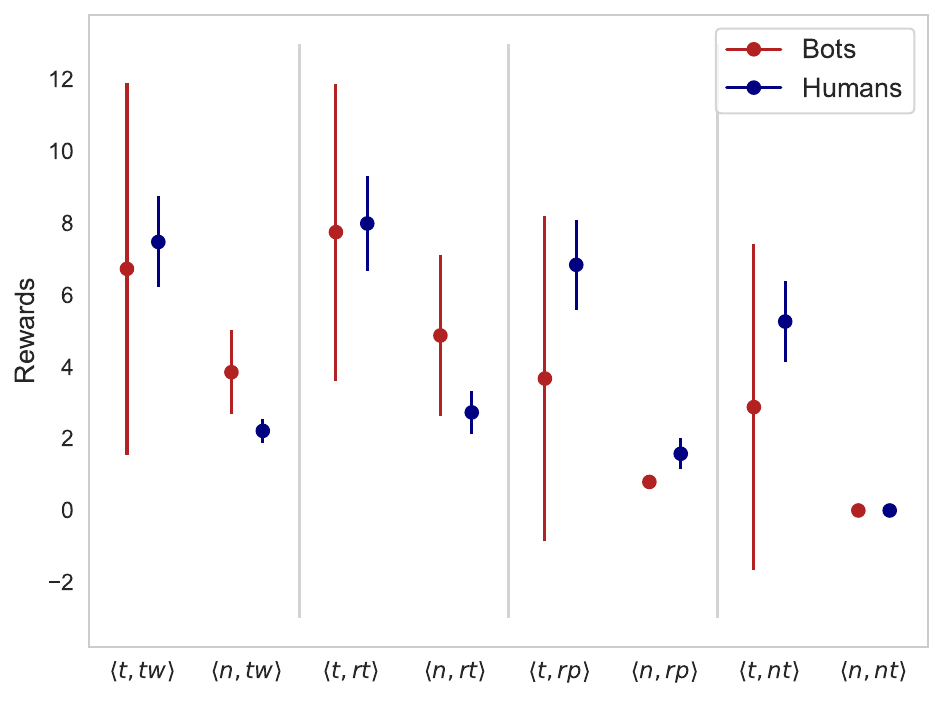}
            \caption{States tweets and nothing}\label{fig:2a}
          \end{subfigure}
          \begin{subfigure}[t]{0.49\textwidth}
            \centering
            \includegraphics[width=\textwidth]{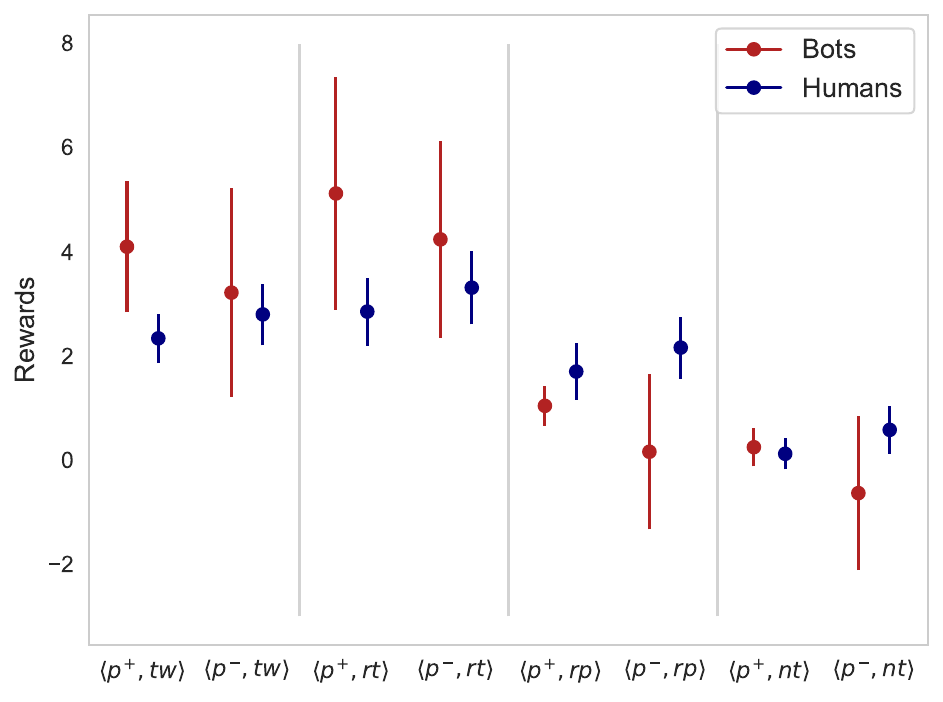}
            \caption{States supporting and opposing replies}\label{fig:2b}
          \end{subfigure}

    \caption{Means and standard errors of the estimated rewards for the 16 {state-action} pairs for bots and humans. }
    \label{fig:mean_err_IRL}
\end{figure}

After analyzing the overall strategy of propagandists, we classify our data into bots and humans to account for differences in the strategy by the two distinct groups. Overall, 11.05\% of the accounts are classified as bots and 88.95\% as humans. Figure~\ref{fig:mean_err_IRL} shows the mean rewards (with standard errors) for bots and humans for each \emph{state-action} pair $\langle s, a \rangle$, $s \in \mathcal{S}$ and $a \in \mathcal{A}$. 

The main strategy of humans is to perform an action (whether that be a reply, retweet, or source tweet) when having been retweeted. This indicates that humans strategically make use of the resharing behavior of other users to disseminate propaganda. Similarly, bots tend to perform an action after having been retweeted. In particular, bots are likely to post source tweets and retweets. Hence, also bots make strategic use of the resharing behavior of others, in particular, to spread new content.

\begin{figure*}[htp]
  \centering
  \vspace{-0.4cm}
      \begin{subfigure}[t]{0.49\textwidth}
        \centering
        \includegraphics[width=\textwidth]{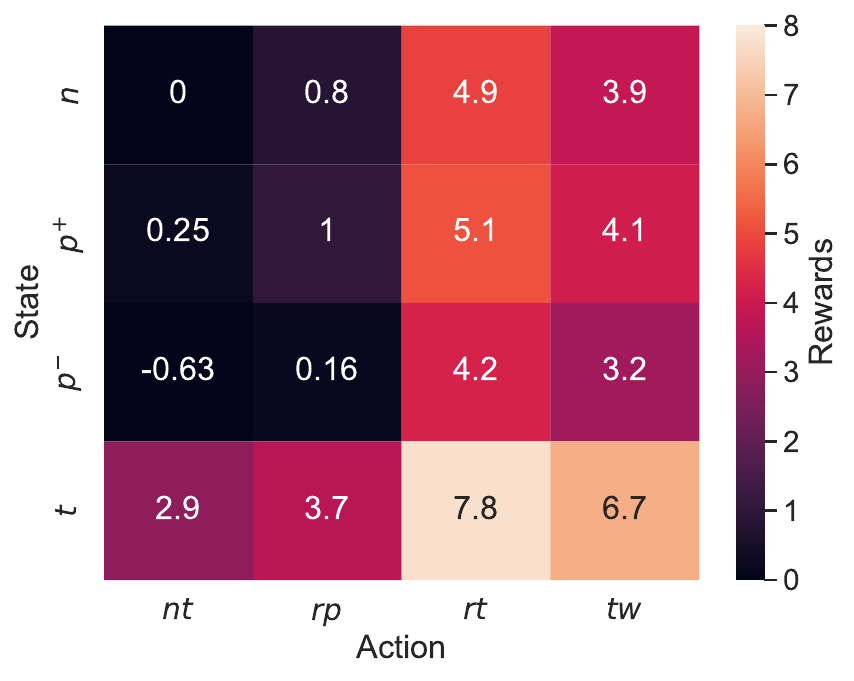}
        \caption{Bots}\label{fig:3a}
      \end{subfigure}
      \begin{subfigure}[t]{0.49\textwidth}
        \centering
        \includegraphics[width=\textwidth]{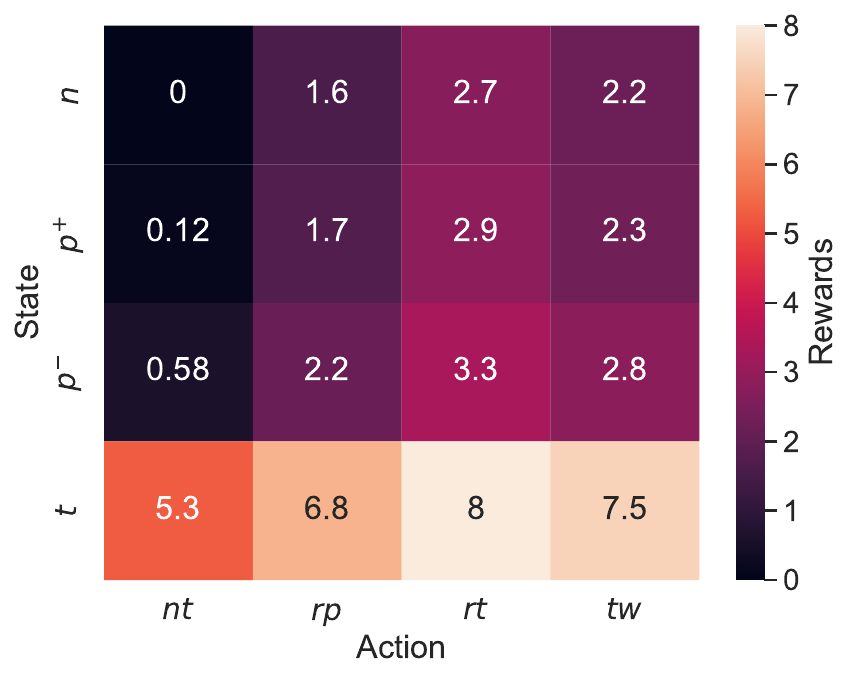}
        \caption{Humans}\label{fig:3b}
      \end{subfigure}
      \caption{Average estimated rewards for {state-action} pairs for bots and humans.}
      \label{fig:heatmap}
\end{figure*}

Further, we compute the average rewards for each {state-action} pair for (a)~bots and (b)~humans, which corroborate our above findings (Figure~\ref{fig:heatmap}). It can be seen that humans and bots have high rewards to take an action when they have been retweeted ($s = \mathsf{t}$). At the same time, bots have high rewards to post source tweets and retweets ($a = \mathsf{tw,rt}$) in all states. Hence, for humans, the resharing behavior is an important driver for the dissemination strategy of propaganda, while bots do not only rely on reshares. Instead, bots also aim to spread propaganda by sharing organically appearing propaganda when receiving replies or no interactions.

Next, we compare the rewards for bots and humans after receiving a supportive or opposing reply or mention. Overall, bots are more incentivized to act when receiving a supportive reply in comparison to an opposing reply. In contrast, humans are more incentivized through opposing replies. This might point to a network of bots that support each other through replies, while humans are more likely to engage also in critical discussions.

\begin{figure}
\centering
\includegraphics[width=0.5\textwidth]{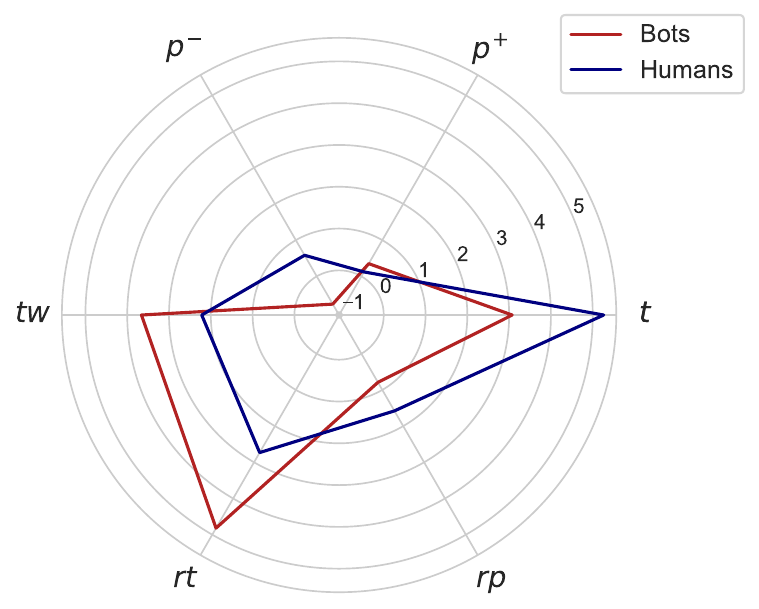}
\caption{Weights of the reward function in our IRL algorithm. Shown is a breakdown by the states and actions for bots and humans.}
\label{fig:spiderchart_IRL}
\end{figure}

As an additional check, we also compute the average weights of each state and action for humans and bots as a measure of feature importance (see Figure~\ref{fig:spiderchart_IRL}). Our findings confirm what we found previously: bots make use of posting new content or resharing content of others to disseminate propaganda while humans rely most on being reweeted by others in the network. The importance of supportive replies is similar for bots and humans, while negative replies are a more important driver for humans.

\subsection{Robustness checks}
\zlabel{sec:robustness_checks}

We performed a series of robustness checks, which corroborated the above findings: 
\begin{enumerate}
\item We validated the bot classification using a second supervised machine learning algorithm, namely Bot Sentinel\footnote{\url{https://botsentinel.com}}. Bot Sentinel focuses on classifying inappropriate accounts on Twitter, which includes bots, trolls, and coordinated accounts. The validation of a subset of 2,661 accounts revealed an agreement of 61.93\% between Botometer and Bot Sentinel (which can be explained by the slightly different definition of how classifications should be made). Yet, the proportion of bots in the validation set was very similar for both classifiers. 
\item We validated the IRL results using the bot detection from Bot Sentinel. Here, the estimated mean rewards for the {state-action} pairs of humans are similar to that of our main analysis. Both show large rewards for all actions in state $\mathsf{t}$. For bots, we find higher rewards for retweeting (action $\mathsf{rt}$), while for humans rewards for replies are slightly larger. Howevere, slight differences in the results are expected due to the slightly different label definition.
\item We performed sensitivity analyses for maximum entropy IRL. Analogous to \citet{Luceri.2020}, the trajectory was set to have a minimum length of five.  Nevertheless, we also experimented with other lengths, but found similar results. Generally, shorter lengths tend to suffer from instability during estimation and thus have a larger variance, which is why we set the default to five. 
\item We also verified our stance detection with more strict thresholds for the binary classification. We classified replies and mentions as supportive when the average predicted stance was $> 0.7$ and as opposing when the average predicted stance was $< 0.3$. The robustness check corroborates our previous findings that humans rely upon the resharing behavior of others to disseminate propaganda. In return, bots are more likely to retweet and post new content in all states, particularly when having been retweeted. We also corroborate that bots are more incentivized by positive replies and mentions.
\item We further experimented with different discount factors $\gamma$, but again arrived at conclusive findings.
\end{enumerate}

\section{Discussion}
\label{sec:discussion}

\textbf{Relevance:} 
Social media is widely used for information warfare \cite{Bail.2020, Doroshenko.2021, Ferrara.2017}. A prominent example is the proliferation of propaganda during the 2022 Russian invasion of Ukraine. The proliferation of propaganda during this critical event underscores the urgent need to understand the strategy behind the campaign. To the best of our knowledge, this is the first study that leverages IRL to analyze the strategy of propagandists when shaping the online discourse on the 2022 invasion. 

Our methodological approach offers distinct advantages over simple descriptive analyses. A unique strength of IRL is that it allows us to model the actual sequence of online interactions over time and thus account for interdependencies in how propagandists influence and are influenced by other social media users. Particularly, we can discover how propagandists react to support and opposition in online communities. Understanding the underlying strategy employed by propagandists is a crucial step towards identifying and ultimately curbing the spread of coordinated propaganda campaigns on social media platforms. 

\textbf{Interpretation:} 
Our findings contribute to the existing research on Russian propaganda on social media by quantifying the strategies with which between-user interactions are strategically used to shape the online discourse. We find that users strategically make use of the resharing behaviors of others while they mostly post new content and reshare others. When classifying the data into bots and humans, we find that humans are more likely to react to opposing replies, and that opposing replies are a more important driver for humans according to the feature importance. We interpret this as an indication that humans are more likely to engage in critical discussions with other users. By responding to opposing perspectives, human users demonstrate a propensity for critical engagement. Conversely, our analysis highlights that bots are more incentivized by supporting replies. This observation suggests that bots may predominantly focus on reinforcing pre-existing narratives and contribute to the formation of support networks.

\textbf{Comparison to literature:} Previous work has examined the scope and actors of Russian propaganda on social media during the 2022 invasion of Ukraine \citep{Pierri.2022,Geissler.2022}, yet the strategy with which propagandists' use between-user interactions to shape the online discourse has remained unclear. Bots and humans are likely working in coordination since they aim to achieve a common goal. In this regard, collective action theory argues that members of a group work together to achieve a common goal \cite{Olson.1965}. Different from descriptive analyses, we leverage IRL to explicitly model the \emph{sequence} of online interactions over time, which allows us to assess the role of between-user interactions and thus shed light on the coordination tactics behind Russian propaganda.

Our findings on the dissemination strategy of bots are also different from findings on the spreading of low-credibility content \cite[e.g.,][]{Shao.2018}. We find that bots mostly rely on retweeting to spread online propaganda, instead of humans being the main retweeters. Moreover, previous work found that a key strategy for bots is to spread content by mentioning influential users \cite{Shao.2018}. In contrast, we find that bots do not use  replies or mentions to spread online propaganda. Instead, they are incentivized by support from the social community. A possible explanation for this is that bots aim to increase the traffic around pro-Russian propaganda, which can lead to certain hashtags appearing as ``trending topics'' on the frontpage of Twitter and thus being visible to all users. This is supported by anecdotal evidence in that the media even commented on the observation that pro-Russian propaganda was shown under Twitter's ``trending topics'' \cite{BBC.2022}.

\textbf{Implications:} Our results have direct implications for social media platforms and societies. Our research contributes to the broader understanding of information warfare and its implications for online discourse and societal well-being. We highlight the vulnerabilities of social media to be manipulated for information warfare. Here, understanding the strategy of propagandists can be used for surveillance and early warning systems to detect coordinated propaganda campaigns on social media. For example, the reward structure can be used as input for machine learning systems to detect and subsequently remove activities that amplify the spreading of propaganda. Another promising way to curb the impact of online propaganda is to educate users on how to recognize and handle propagandists in online interactions so that humans can identify when propaganda is shared and amplified. In this regard, it may also be likely that behavioral interventions such as counter-measures against fake news \cite{Gallotti.2020, Kozyreva.2022, Pennycook.2021, Garrett.2013} can be adapted to online propaganda. However, such interventions mainly target individual users, and, as our findings imply, their application to propaganda campaigns might be limited when the campaigns leverage between-user interactions strategically and thus show characteristics of coordination efforts.

A concerning implication of our work is that, without significant interventions from social media platforms, propaganda campaigns can strategically use between-user interaction to shape online narratives, which can have real-world consequences. Propaganda on social media can influence public opinion on political and military matters and increase division in society. These risks can become larger and harder to manage through the amplification by bots. This is especially alarming since repeated exposure lets people believe false information, even when they know it is false \cite{Pennycook.2018}. It is thus important that policy-makers are aware of the potential threats that social media propaganda poses to modern societies.

\textbf{Limitations and future research:} As with other research, ours is not free of limitations that offer opportunities for future research. First, our research is based on only one social media platform. However, Twitter is a platform with a particularly large and international audience\footnote{\url{https://www.statista.com/statistics/242606/number-of-active-twitter-users-in-selected-countries/}} and is a common source for news consumption\footnote{\url{https: //www.pewresearch.org/journalism/2021/11/15/news-on-twitter-consumed-by-most-users-and-trusted-by-many/}}. Hence, it poses a fertile ground for propaganda. Moreover, future research could extend our analysis to alt-right social media platforms \cite{Bar.2023b}. Second, we focus on the propaganda narrative from the beginning of the 2022 Russian invasion of Ukraine. Previous research shows that the main propaganda activity happened right before and in the beginning of the invasion \cite{Geissler.2022, Pierri.2022}. Nevertheless, future work may want to repeat our analysis to account for that propaganda campaigns tend to get more refined over time. Third, the accuracy of our analysis depends on the accuracy of other tools such as Botometer. However, these tools have shown to have high accuracy in detecting bot activity \cite{Varol.2017} and are widely used in research \cite{Ferrara.2017, Shao.2018}.

\section{Conclusion}
\label{sec:conlusion}
While the scope and actors of the Russian propaganda campaign from the invasion of Ukraine have been researched, little is known about the strategy of bots and humans when disseminating propaganda. To fill this gap, we model the online behavior of propagandists over time via a MDP, where we carefully account for the sequential dynamics of between-user interactions. Using IRL, we then estimate the underlying reward structure, which characterizes the dissemination strategy of bots and humans. We find that bots are more likely to post source tweets and retweet other users after having been retweeted. Humans, in return, strategically elicit resharing behavior of others to make their content go viral. Moreover, bots rely on support from the online community while humans are more likely to react to opposing replies. Altogether, our findings help to understand the dissemination strategy of propaganda and, hence, can aid in the prevention of future propaganda campaigns on social media.

\section*{Author contributions}
All authors contributed to conceptualization, results interpretation, and manuscript writing. First author contributed to data analysis. All authors approved the manuscript.


\bibliography{literature.bib}

\clearpage
\appendix

\vspace{1cm}
\begin{center}
\huge Supplementary Materials
\end{center}
\vspace{1cm}

\section{Overview of robustness checks} \label{supp:robustness_checks} 
\noindent 
We conducted an extensive set of checks and complementary analyses to validate the robustness of the bot detection, stance classification, and alternative model parameters. We briefly present the main findings in the following.

\subsection{Robustness of the bot detection}
In our main analysis, we classify the user accounts into bots and humans using Botometer (11.05\% bots and 88.95\% humans). We validated the bot classification with Bot Sentinel, a supervised machine learning classifier that identifies inappropriate accounts. Bot Sentinel requires a minimum of 10 messages per account to make a classification. A total of 2,661 accounts in our dataset complied with this requirement. Overall, the agreement between the two classifiers was 61.93\% while the proportion of bot accounts remained consistent across the classifiers. We explain this difference in agreement by the slightly different definition of accounts to flag: Botometer is specifically designed to detect automated accounts (bots) \cite{Varol.2017}, whereas Bot Sentinel detects inappropriate accounts, which can be bots, trolls, or other spamming accounts.\footnote{\url{https://botsentinel.com}}

We repeated the IRL with the accounts labeled as bots and humans by Bot Sentinel and found similar results to the main analysis (see Supplementary Figure~\ref{fig:supp_heatmap_val}). Both bots and humans strategically make use of the resharing behavior of others to shape online discourse dynamics. Further, bots are incentivized to retweet others whereas humans also make use of the reply function.

\begin{figure*}[h]
  \centering
  \vspace{-0.4cm}
      \begin{subfigure}[t]{0.49\textwidth}
        \centering
        \includegraphics[width=\textwidth]{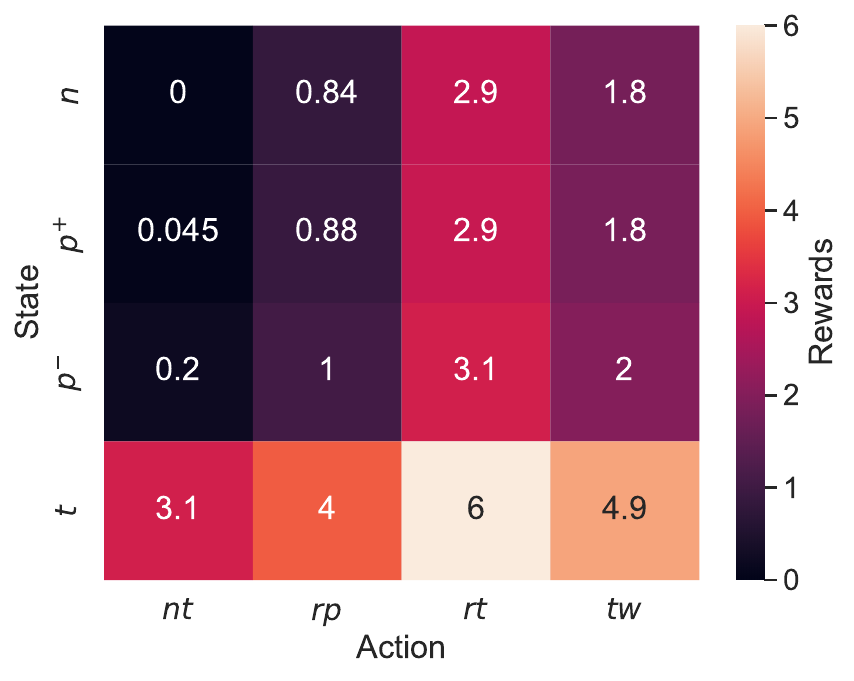}
        \caption{Bots}\label{fig:3a_val}
      \end{subfigure}
      \begin{subfigure}[t]{0.49\textwidth}
        \centering
        \includegraphics[width=\textwidth]{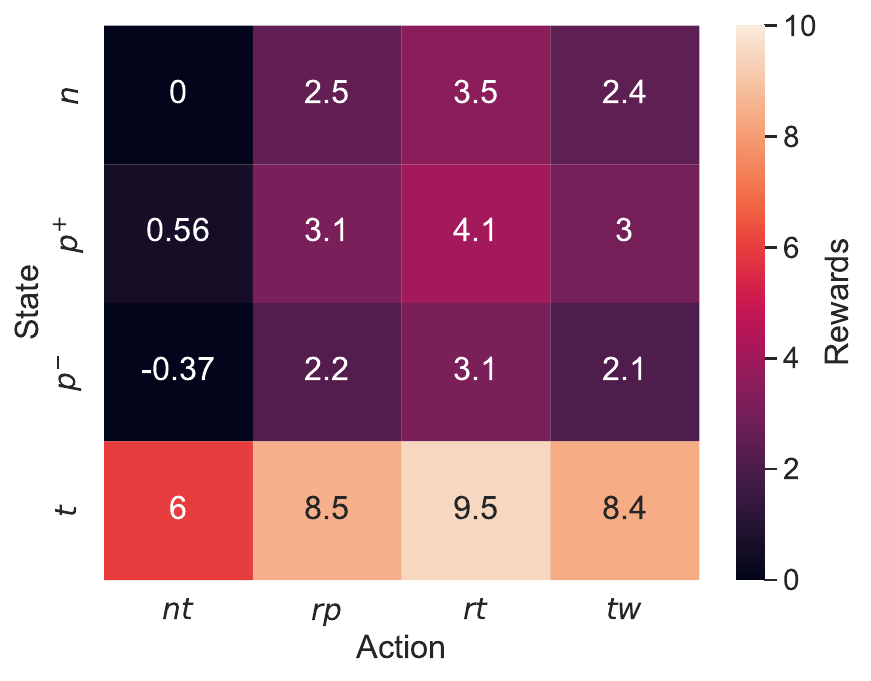}
        \caption{Humans}\label{fig:3b_val}
      \end{subfigure}
      \caption{Average estimated rewards for {state-action} pairs for bots and humans labelled by Bot Sentinel.}
      \label{fig:supp_heatmap_val}
\end{figure*}

\subsection{Robustness of stance classification}
\begin{figure*}[h]
  \centering
  \vspace{-0.4cm}
      \begin{subfigure}[t]{0.49\textwidth}
        \centering
        \includegraphics[width=\textwidth]{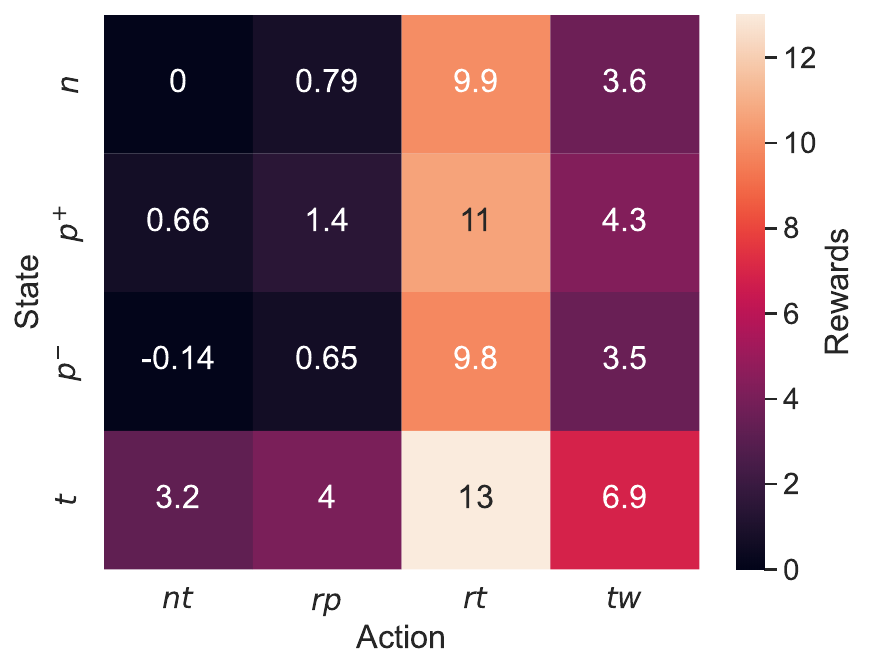}
        \caption{Bots}\label{fig:3a_robustness}
      \end{subfigure}
      \begin{subfigure}[t]{0.49\textwidth}
        \centering
        \includegraphics[width=\textwidth]{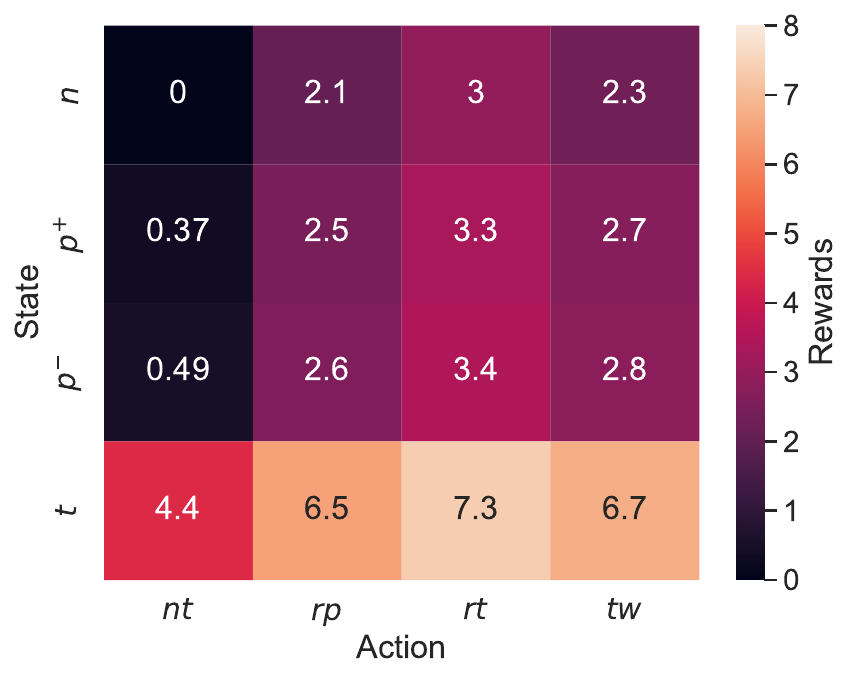}
        \caption{Humans}\label{fig:3b_robustness}
      \end{subfigure}
      \caption{Robustness check: Average estimated rewards for {state-action} pairs for bots and humans with clear stance messages.}
      \label{fig:supp_heatmap_robustness}
\end{figure*}

We further evaluated the robustness of the results regarding the stance classification of replies and mentions. For this, we applied more strict thresholds for the replies and mentions labeled by BERTweet. We labeled messages as supporting when the predicted stance was $> 0.7$ and as opposing when the predicted stance was $< 0.3$. Supplementary Figure~\ref{fig:supp_heatmap_robustness} shows the mean rewards for state-action pair for (a)~bots and (b)~humans. The results of the robustness check provide further support for our previous findings, affirming that humans heavily depend on the resharing behavior of others to propagate propaganda. In contrast, bots exhibit a greater inclination to retweet and post new content across all states, but particularly when they have been retweeted themselves. Additionally, our analysis reinforces the observation that bots show a stronger preference for positive replies and mentions as incentives for their engagement. These findings enhance our understanding of the distinct behaviors exhibited by humans and bots in online discourse, highlighting the reliance of humans on social reinforcement through resharing while emphasizing the bots' responsiveness to positive feedback.

\subsection{Robustness of IRL parameters}
To validate the maximum entropy IRL, we conducted sensitivity analyses for the IRL algorithm, specifically focusing on the minimum length of the trajectory. To explore the impact of trajectory length on our results, we systematically varied the parameter between 3, 5, and 10. We observed that shorter trajectory lengths were more susceptible to estimation instability, leading to higher variance in the results. On the other hand, longer trajectory lengths decreased the number of data points drastically. This leads to less generalizable results and a potential bias for particularly active users. Further, we varied the discount factor between 0.7 and 0.9 but found consistent results.

\end{document}